\documentclass[a4paper,11pt]{article}
\pdfoutput=1

\usepackage{jheppub} 
\usepackage[T1]{fontenc}
\usepackage[utf8]{inputenc}
\usepackage{amsmath}
\usepackage{amsfonts}
\usepackage{amssymb}
\usepackage{latexsym}
\usepackage{mathrsfs}
\usepackage{braket}		
\usepackage{twistor}

\newcommand{\msf}[1]{\mathsf{#1}}
\newcommand{\sC}{\mathsf{C}}
\newcommand{\spl}{\text{split}}
\newcommand{\ope}{\text{ope}}
\newcommand{\CS}{\mathcal{CS}}

\title{Descendants in celestial CFT and emergent multi-collinear factorization}

\author[a]{Stephen Ebert,}
\author[b]{Atul Sharma}
\author[c]{and Diandian Wang}

\affiliation[a]{Mani L. Bhaumik Institute for Theoretical Physics,\\
University of California, Los Angeles, CA 90095, USA}
\affiliation[b]{The Mathematical Institute,\\ University of Oxford, Woodstock Road, OX2 6GG, United Kingdom}
\affiliation[c]{Department of Physics,\\ University of California, Santa Barbara, CA 93106, USA}

\emailAdd{stephenebert@physics.ucla.edu}
\emailAdd{atul.sharma@maths.ox.ac.uk}
\emailAdd{diandian@physics.ucsb.edu}

\abstract{Multi-collinear factorization limits provide a window to study how locality and unitarity of scattering amplitudes can emerge dynamically from celestial CFT, the conjectured holographic dual to gauge and gravitational theories in flat space. To this end, we first use asymptotic symmetries to commence a systematic study of conformal and Kac-Moody descendants in the OPE of celestial gluons. Recursive application of these OPEs then equips us with a novel holographic method of computing the multi-collinear limits of gluon amplitudes. We perform this computation for some of the simplest helicity assignments of the collinear particles. The prediction from the OPE matches with Mellin transforms of the expressions in the literature to all orders in conformal descendants. In a similar vein, we conclude by studying multi-collinear limits of graviton amplitudes in the leading approximation of sequential double-collinear limits, again finding a consistency check against the leading order OPE of celestial gravitons.}

\begin{document} 
\maketitle
\flushbottom

\section{Introduction}
\label{sec:intro}

Recent decades have seen many discoveries of alternative mathematical structures from which the standard principles of perturbative QFT emerge as derived consequences. One of the primary motivations of such investigations has been to find a holographic description for scattering amplitudes in flat space, akin to the highly successful AdS/CFT paradigm. In this context, celestial conformal field theory (CCFT) is a recent proposal that claims to identify Yang-Mills and gravitational amplitudes in $\R^{1,3}$ with correlators of a putative $2d$ CFT living on the celestial sphere at null infinity. And even though no explicit candidate or stringy construction for such a holographic dual has been found yet, great progress is being made in understanding the abstract structures and symmetries that such a CFT could possess. Some of the notable advances include the main work on celestial amplitudes \cite{PSS,Pasterski:2017kqt,Pasterski:2017ylz,Lam:2017ofc,Schreiber:2017jsr,Stieberger:2018edy,Nandan:2019jas,Law:2019glh,Albayrak:2020saa,Casali:2020vuy,Law:2020xcf}, on asymptotic symmetries and soft theorems \cite{Strominger:2013lka,Strominger:2013jfa,StromingerNotes,Cheung:2016iub,Donnay:2018neh,Stieberger:2018onx,Banerjee:2018gce,Pate:2019mfs,Adamo:2019ipt,Puhm:2019zbl,Guevara:2019ypd,Fotopoulos:2019tpe,Fotopoulos:2019vac,Donnay:2020guq,Fan:2020xjj,Fotopoulos:2020bqj}, and on the CCFT operator algebra \cite{Fan:2019emx,Pate:2019lpp,Banerjee:2020kaa,Banerjee:2020zlg}.

Due to the absence of an actual candidate CFT, most of the work has been kinematical and one-sided: trying to understand the CFT side by studying properties of the amplitudes. This begs the question: how do we begin to \emph{test} this holographic proposal? An interesting direction was taken in \cite{Pate:2019lpp}, which provided a purely holographic derivation of the CCFT operator product expansions (OPE) via imposing asymptotic symmetry constraints. In turn, this gave a new holographic foundation for the universality of the well-known collinear limits of gluon and graviton amplitudes \cite{Altarelli:1977zs,Bern:1998sv,Taylor:2017sph}. The work of \cite{Banerjee:2020kaa,Banerjee:2020zlg} took this further by showing that even subleading terms in the collinear expansions of low-multiplicity graviton amplitudes can be ascribed to BMS descendants in the gravitational CCFT. Coupled with the idea that a CFT is in principle completely determined from its CFT data, i.e., its operator content and OPE algebra, such computations are enough to allow us to come up with some simple tests of the duality.

One of the hallmarks of scattering amplitudes is the structure of their factorization poles and residues. These are completely fixed by the principles of locality and unitarity. An important test of any holographic dual would then be to discover them as emergent properties of the corresponding CFT correlators. From the viewpoint of the OPE, the most natural object to study in this regard are multi-collinear limits of the amplitudes. These are maximally singular limits that recursively probe all possible factorization poles and residues (see \cite{Birthwright:2005ak} and references therein). In this work, we show that these can indeed be holographically determined by the symmetries and OPE of the dual CCFTs. This provides an example of a calculation that utilizes the CCFT to essentially ``bootstrap'' the physics of amplitudes.

Such a reconstruction of the bulk physics requires us to understand the contributions of descendants to the celestial OPE, and we will mostly focus on the gluon OPE for sake of simplicity. After a review of some standard material in \S\ref{sec:background} and \S\ref{sec:algebra}, we begin with this task in \S\ref{sec:conformal}. Global supertranslation symmetry is used to fix the OPE coefficients of all the (global) conformal descendants of celestial gluon operators. For completeness, in \S\ref{sec:km} we also compute examples of Kac-Moody descendants contributing to the OPE of two positive helicity gluons. This is done by imposing Poincar\'e as well as Kac-Moody invariance. In fact, naively the symmetries overdetermine the OPE coefficients, but the results are beautifully mutually consistent. In appendix \ref{app:KM}, we directly verify that these descendants are exchanged in the 4-gluon amplitude with precisely the predicted OPE coefficients.

In \S\ref{sec:gluon}, we look at multi-gluon collinear limits. In the language of conformal correlators, this corresponds to bringing multiple operators close together. To get a feel for the idea, consider a correlator of $N$ operators in a CFT$_2$,
\be\label{cftcor}
\left\la O_1(z_1,\bar z_1)\,O_2(z_2,\bar z_2)\cdots O_N(z_N,\bar z_N)\right\ra\,,
\ee
with $(z_i,\bar z_i)$ denoting complex coordinates. The various operators have operator products taking the generic form,
\be\label{cftope}
O_i(z_i,\bar z_i)\,O_j(z_j,\bar z_j) \sim \sum_k \sC_{ijk}\,O_k(z_j,\bar z_j)\,,
\ee
with $\sC_{ijk}\equiv \sC_{ijk}(z_{ij},\bar z_{ij},\p_j,\dbar_j)$ denoting Wilson coefficients that depend on the CFT data and $z_{ij} \equiv z_i - z_j$. Now, for operators in a given ordering, say $|z_{12}|<|z_{23}|<\cdots<|z_{n-1,n}|$ (where $n\leq N$), we can replace the $n$-fold product $\prod_{i=1}^nO_i(z_i,\bar z_i)$ by using the OPE to perform $n$ sequential Wick contractions:
\be\label{opetocor}
\sum_{\{k_a\}}\sC_{12k_1}\sC_{k_13k_2}\cdots \sC_{k_{n-2},n,k_{n-1}}\left\la O_{k_{n-1}}(z_n,\bar z_n)\,O_{n+1}(z_{n+1},\bar z_{n+1})\cdots O_N(z_N,\bar z_N)\right\ra\,.
\ee
Then the product of Wilson coefficients gives the equivalent of a multi-collinear splitting function (or more appropriately ``splitting operator'') for CFT correlators. It is automatically universal since it does not depend on the other $N-n$ operators inserted in the correlator.

Thus, having determined the CCFT Wilson coefficients to the desired accuracy, we can approximate celestial amplitudes with such recursive celestial OPEs in the limit of small $z_{ij}$. As the main utility of these coefficients, we will find the leading multi-collinear factorization behavior of the usual momentum space amplitudes without any input whatsoever from Feynman rules or the usual techniques of 4$d$ QFT. Finally, the existence of a CCFT interpretation can guarantee the universality of these limits. Some integration techniques relevant to these computations are described in appendix \ref{app:integrals}.

In \S\ref{sec:graviton}, we attempt a similar calculation for graviton amplitudes. Due to a lack of literature to compare with on the gravitational side, we will only be able to outline a leading order computation for the simplest multi-graviton collinear limits. This nevertheless provides a nice consistency check of the formalism and a motivation for further work.

\paragraph{Note added:} Since this paper was completed, new work \cite{Banerjee:2020vnt} has shown that one also finds descendants of subleading soft gluon symmetries in the celestial gluon OPE. We have neglected these here but suspect that they will be required for a more complete derivation of factorization limits in the future. 


\section{Background}\label{sec:background}

In this section, we collect some standard conventions about celestial amplitudes and results for celestial gluon and graviton operator product expansions. Then we review the basics of multi-collinear limits that will come to use later, noting relevant results from the literature.


\subsection{Celestial amplitudes and OPE}

The null 4-momentum $k_{\al\dal}$ of a typical massless particle can be decomposed as
\be\label{kdef}
k_{\al\dal} = s\,\omega\,q_{\al\dal}(z,\bar z)\,,\qquad q_{\al\dal}(z,\bar z) = \begin{pmatrix}1 &\; \bar z\\z &\; z\bar z\end{pmatrix}_{\al\dal}\,.
\ee
Here, the sign $s=\pm 1$ denotes whether the particle is outgoing or incoming, while $\omega\in\R^+$ denotes its energy. The remaining null vector $q_{\al\dal}$ stands for the embedding of the celestial sphere $\CS^2$ as the projective light cone of any point in flat space, with $z,\bar z$ giving complex coordinates on the sphere. By convention, the corresponding spinor-helicity variables are taken to be
\be\label{shdef}
k_{\al\dal} = \lambda_\al\,\bar\lambda_{\dal}\,,\qquad\lambda_\al = \sqrt{\omega}\begin{pmatrix}1\\z\end{pmatrix}_\al\,,\quad\bar\lambda_{\dal} = s\sqrt{\omega}\begin{pmatrix}1\\\bar z\end{pmatrix}_{\dal}\,.
\ee
With Lorentzian signature $(-\,+\,+\;+)$, one chooses the standard reality condition on these: $\bar z = z^*$ (complex conjugation). We will stick to this, except for using split signature $(-\,-\,+\;+)$ in appendix \ref{app:KM}.

``Celestial amplitudes'' are the scattering amplitudes of conformal primary wavepackets of gluons and gravitons. In short, they can be defined in terms of a change of basis implemented by Mellin transforms:
\be\label{celamp}
\cA_n(\{s_i,\Delta_i,\ell_i,z_i,\bar z_i\}) = \prod_{j=1}^n\int_0^\infty\frac{\d\omega_j}{\omega_j}\,\omega_j^{\Delta_j}\;A_n(\{k_i,\ell_i\})\;\delta^4\!\left(\sum_{l=1}^nk_l\right)\,.
\ee
Here $i=1,\dots,n$ are particle labels, with $k_i = s_i\,\omega_i\,q_i$, $q_i\equiv q(z_i,\bar z_i)$ as described above. $A_n$ denotes the usual momentum space amplitude. For gluons, it will be further augmented with color indices. $\ell_i=\pm1$ or $\pm 2$ are the gluon/graviton helicities. 

Under a M\"obius transformation of $\CS^2$ coordinates, the celestial amplitudes $\cA_n$ transform conformally covariantly with weights \cite{Pasterski:2017kqt,Pasterski:2017ylz}
\be\label{weights}
h_i = \frac{\Delta_i+\ell_i}{2}\,,\qquad\bar h_i = \frac{\Delta_i-\ell_i}{2}
\ee
in the $(z_i,\bar z_i)$. Consequently, these are conjectured to be the correlators of certain conformal primary operators in a 2$d$ CFT living on $\CS^2$, called a celestial CFT. Such conformal primaries dual to gluons and gravitons are referred to as celestial gluon/graviton operators. Celestial gluons are denoted by $O^{\ell\,\msf{a},\,s}_{\Delta}(z,\bar z)$, where $\msf{a}$ is an adjoint index. Celestial gravitons are commonly denoted by $G^{\ell,s}_\Delta(z,\bar z)$. For most of what follows, we will focus only on outgoing particles for which $s=+1$, so we drop this superscript to avoid cluttering notation.

Collinear singularities $z_{ij} \to 0$ in momentum space amplitudes are interpreted as OPE singularities of the correlators obtained from the Mellin transform. This has allowed a determination of the leading celestial OPE of gluons and gravitons \cite{Fan:2019emx,Pate:2019lpp}. For instance, the OPEs of two outgoing gluons read
\begin{align}
    O^{+\,\msf{a}}_{\Delta_1}(z_1,\bar z_1)\,O^{+\,\msf{b}}_{\Delta_2}(z_2,\bar z_2) &\sim -\frac{\im\,f^\msf{abc}}{z_{12}}\,B(\Delta_1-1,\Delta_2-1)\,O^{+\,\msf{c}}_{\Delta_1+\Delta_2-1}(z_2,\bar z_2)\,,\label{O+O+}\\
    O^{-\,\msf{a}}_{\Delta_1}(z_1,\bar z_1)\,O^{+\,\msf{b}}_{\Delta_2}(z_2,\bar z_2) &\sim -\frac{\im\,f^\msf{abc}}{z_{12}}\,B(\Delta_1+1,\Delta_2-1)\,O^{-\,\msf{c}}_{\Delta_1+\Delta_2-1}(z_2,\bar z_2)\nonumber\\
    &\qquad\qquad-\frac{\im\,f^\msf{abc}}{\bar z_{12}}\,B(\Delta_1-1,\Delta_2+1)\,O^{+\,\msf{c}}_{\Delta_1+\Delta_2-1}(z_2,\bar z_2)\label{O-O+}\,,
\end{align}
and similarly for the $O^-\,O^-$ case, with $f^\msf{abc}$ denoting the gauge group's structure constants and $B(a,b)$ referring to the Euler beta function. For gravitons, one instead finds
\begin{align}
    G^+_{\Delta_1}(z_1,\bar z_1)\,G^+_{\Delta_2}(z_2,\bar z_2) &\sim -\frac{\kappa}{2}\,\frac{\bar z_{12}}{z_{12}}\,B(\Delta_1-1,\Delta_2-1)\,G^+_{\Delta_1+\Delta_2}(z_2,\bar z_2)\,,\label{grav2ope++}\\
    G^-_{\Delta_1}(z_1,\bar z_1)\,G^+_{\Delta_2}(z_2,\bar z_2) &\sim -\frac{\kappa}{2}\,\frac{\bar z_{12}}{z_{12}}\,B(\Delta_1+3,\Delta_2-1)\,G^-_{\Delta_1+\Delta_2}(z_2,\bar z_2)\nonumber\\
    &\qquad\qquad-\frac{\kappa}{2}\,\frac{z_{12}}{\bar z_{12}}\,B(\Delta_1-1,\Delta_2+3)\,G^+_{\Delta_1+\Delta_2}(z_2,\bar z_2)\,,\label{grav2ope-+}
\end{align}
where $\kappa =\sqrt{32\pi G_N}$ is the gravitational coupling.


\subsection{Multi-collinear factorization}

Gluon and graviton amplitudes in flat space are completely characterized by on-shell methods like BCFW recursion and MHV diagrams \cite{Cachazo:2004kj,Britto:2004ap,Britto:2005fq}. Such relations follow from their factorization poles and residues as a consequence of locality and unitarity. However, unlike collinear limits, general factorization limits don't seem to have an obvious analog for celestial amplitudes. One possibility to make a connection between the two formalisms is thus to study more involved collinear singularities, namely multi-collinear limits. They correspond to simultaneously taking all possible factorization limits involving a certain subset of the scattering particles. Such limits have been well-studied in the literature on QCD, with important progress originating from the usage of CSW rules and MHV diagrams \cite{Birthwright:2005ak}. For gravity, there has been no such progress beyond the 2-graviton collinear limit, more or less due to a lack of strong theoretical foundations for similar recursive methods.\footnote{Gravitational MHV rules seem to work only for amplitudes involving $n\leq11$ gravitons \cite{Bianchi:2008pu}.}

The kinematic configuration probing a multi-collinear singularity corresponds to the null momenta $k_1,\dots,k_n$ of a subset of the particles becoming collinear. For simplicity, we take all of these to be outgoing. Then all the propagators of the form $(k_{i_1}+\cdots+k_{i_r})^{-2}$, $i_1,\dots,i_r\in\{1,\dots,n\}$, diverge. This leads to a maximally singular sub-amplitude to bubble off, yielding a universal factor called a splitting function. From a $4d$ perspective, its universality in the gluon case again follows from MHV diagrams. From the holographic viewpoint, our claim is that its universality is a consequence of the celestial OPE --- an argument which might also extend to gravity.

To make the limits precise and set up some notation, note that we can always express the sum of multiple momenta in terms of two auxiliary null momenta,
\be\label{ksum}
k_1+k_2+\cdots+k_n = p + \epsilon\,n\,,
\ee
where for instance we can somewhat canonically choose $n$ to be the null generator of $\scri^+$. It follows that since $\omega_i=n\cdot k_i$, this choice results in
\be\label{omegap}
\omega_p = \sum_{i=1}^n\omega_i\,,
\ee
where $\omega_p=n\cdot p$ is the energy of $p$. Define the longitudinal-momentum fractions $\xi_i := \omega_i/\omega_p$. The collinear regime is defined by
\be\label{coldef}
k_i \sim \xi_i\,p + \cO(\epsilon)\,,\qquad\forall\;i\in\{1,\dots,n\}\,,
\ee
along with taking $\epsilon$ to be infinitesimal.

In this regime, an $N$-point momentum space amplitude (with $n\leq N$) factorizes as
\be\label{collim}
A_N(1^{\ell_1}\dots N^{\ell_N})\sim \sum_\ell \spl\,(1^{\ell_1}\dots n^{\ell_n}\to p^\ell)\,A_{N-n+1}(p^{\ell}\,(n+1)^{\ell_{n+1}}\dots N^{\ell_N})\,,
\ee
where the superscripts are particle helicities. The universal splitting functions $\spl\,(\cdots)$ are neatly organized by the number of negative helicity gluons participating in the collinearity. If $1,2,\dots,k$ are negative helicity among the $n$ collinear particles, then the corresponding splitting function is denoted by
\be\label{splitdef}
\spl^{(n)}_\ell(1^-\,2^-\dots k^-)\,.
\ee
These will take center-stage in the latter half of this work. 

For gluons, we will content ourselves with considering the $k=0,1$ cases. In this case, the collinear gluons must be adjacent for the splitting function to be maximally singular. For these, the results for the splitting functions come in fairly compact expressions found in \cite{Birthwright:2005ak}. Using the convention \eqref{shdef}, we can write them in variables adapted to the celestial sphere. The simplest of these occur in the case when all the collinear gluons have positive helicity (in this case we denote them by $\spl_\ell^{(n)}$ without any arguments),
\be\label{spl0}
\spl^{(n)}_{-} = 0\,,\qquad\spl^{(n)}_{+} = \frac{\omega_p}{\omega_n \prod^{n-1}_{i = 1} \omega_i \, z_{i,i+1}}\,,
\ee
having stripped off color factors (these will be reinstated for comparison with OPE later). When the first gluon is negative helicity, one finds
\be\label{spl1-}
\spl^{(n)}_{-}(1^-) = \frac{\omega_1^2}{\omega_p\,\omega_n\prod_{i=1}^{n-1}\omega_i \,z_{i,i+1}}\,,
\ee
along with the relatively much more interesting expression,
\begin{multline}\label{spl1+}
    \spl^{(n)}_{+}(1^-) = \frac{\omega_p\,\omega_1^2}{\omega_n\prod_{i=1}^{n-1}\omega_i \,z_{i,i+1}}\Biggl[\sum_{j=2}^{n-1}\frac{(\sum_{l=1}^j\omega_l\,z_{1l})^3}{(\sum_{l=1}^j\omega_l)(\sum_{l=1}^j\omega_l\,z_{jl})(\sum_{l=1}^j\omega_l\,z_{j+1,l})}\,\frac{z_{j,j+1}}{s_{1j}}\\
    \qquad- \frac{(\sum_{l=1}^n\omega_l\,z_{1l})^3}{(\sum_{l=1}^n\omega_l)^2(\sum_{l=1}^n\omega_l\,z_{nl})}\,\frac{1}{s_{1n}}\Biggr]\,,
\end{multline}
where $s_{1j}$ is the generalized Mandelstam variable,
\be\label{mandel}
s_{1j} := \sum_{1\leq k<l\leq j}\omega_k\,\omega_l\,|z_{kl}|^2\,.
\ee
At the level of the first three among these, one only observes two-particle factorization poles. In the language of MHV diagrams, this is a consequence of the fact that only MHV sub-amplitudes happen to blow up for these configurations. The collinearity $1^-\,2^+\cdots n^+\to p^+$ is the first case where NMHV sub-amplitudes can blow up. It will be much more novel for celestial CFT to make contact with multi-particle factorization poles of the sort showing up in \eqref{spl1+}, even if only in leading order approximations in some of the variables. This will be our goal in \S\ref{sec:gluon}.

For gravitons, as mentioned above, there is a distinct lack of data for multi-collinear limits beyond $n=2$. The double-collinear limits that we need are given by \cite{Bern:1998sv}
\be\label{grspl}
\spl^{(2)}_+ = -\frac{\kappa}{2}\,\frac{\bar{z}_{12}}{z_{12}}\, \frac{(\omega_1+\omega_2)^{2}}{\omega_{1}\, \omega_{2}}\,,\qquad\spl^{(2)}_-(1^-) = -\frac{\kappa}{2}\, \frac{\bar{z}_{12}}{z_{12}}\,\frac{\omega_1^3}{\omega_{2}\, (\omega_{1}+\omega_2)^2}\,.
\ee 
The graviton collinear limits are not singular in general and will even depend on the order in which the momenta are made collinear. So we will have to restrict our analysis to the simplest case: using the double-collinear limit to sequentially compute leading order approximations to the multi-collinear splitting functions.


\section{Descendants of celestial gluons}


\subsection{Symmetry algebra}\label{sec:algebra}

In a conformal field theory with an extended symmetry, the states and their corresponding local operators arrange themselves in representations of the symmetry algebra. In the $2d$ CCFT dual to $4d$ Yang-Mills, the conjectured symmetry algebra is Poincar\'e plus a holomorphic Kac-Moody symmetry \cite{He:2015zea,Adamo:2015fwa,McLoughlin:2016uwa}.\footnote{It has level $0$ as far as tree level amplitudes in the bulk are concerned.}${}^{,}$\footnote{One can also instead realize an anti-holomorphic Kac-Moody symmetry, but not both simultaneously, see also \cite{Cheung:2016iub,Fan:2020xjj}.} The representation multiplets are then organized into primaries of the Kac-Moody symmetry and their global conformal descendants and Kac-Moody descendants.

The $4d$ Lorentz group acts as the global conformal group of $\CS^2$. Its generators can be denoted by the standard combinations $\{L_0,\bar L_0,L_{\pm1},\bar L_{\pm1}\}$ of SL$(2,\C)$ dilatations and rotations. The generators of global supertranslations $\cP_{a,b}$ are identified with translation generators (momenta) in the bulk, with the following convenient arrangement:
\be\label{sups}
P_{\al\dal} = \begin{pmatrix}\cP_{-1,-1}&\,\cP_{-1,0}\\\cP_{0,-1}&\,\cP_{0,0}\end{pmatrix}_{\al\dal}\,.
\ee
Their algebra is given by \cite{Stieberger:2018onx,Fotopoulos:2019vac,Banerjee:2020kaa}
\be\label{poincare}
\begin{split}
    &[L_m,L_n] = (m-n)\,L_{m+n}\,,\qquad[\bar L_m,\bar L_n] = (m-n)\,\bar L_{m+n}\,,\\
    &[L_n,\cP_{a,b}] = \left(\frac{n-1}{2}-a\right)\,\cP_{a+n,b}\,,\qquad[\bar L_n,\cP_{a,b}] = \left(\frac{n-1}{2}-b\right)\,\cP_{a,b+n}\,,\\
    &[L_m,\bar L_n] = 0\,,\qquad[\cP_{a,b},\cP_{c,d}] = 0\,.
\end{split}
\ee
Celestial gluons are primaries of the full Poincar\'e group, with transformation laws:
\begin{align}
    &[L_n,O^{\ell\,\msf{a}}_\Delta(z,\bar z)] = \left(z^{n+1}\,\p + (n+1)\,h\,z^n\right)O^{\ell\,\msf{a}}_\Delta(z,\bar z)\,,\label{LO}\\
    &[\bar L_n,O^{\ell\,\msf{a}}_\Delta(z,\bar z)] = \left(\bar z^{n+1}\,\dbar + (n+1)\,\bar h\,\bar z^n\right)O^{\ell\,\msf{a}}_\Delta(z,\bar z)\,,\label{LbarO}\\
    &[\cP_{a,b},O^{\ell\,\msf{a}}_\Delta(z,\bar z)] = z^{a+1}\bar z^{b+1}\,O^{\ell\,\msf{a}}_{\Delta+1}(z,\bar z) \,.\label{PO}
\end{align}
Notice that the action of $\cP_{a,b}$ induces a flow, $(h,\bar h)\mapsto(h+\frac{1}{2},\bar h+\frac{1}{2})$, in the conformal dimensions.

As for the Kac-Moody generators, the holomorphic current is identified with the conformally soft limit of the outgoing positive helicity celestial gluon \cite{Donnay:2018neh,Pate:2019mfs,Fan:2019emx},
\be\label{jdef}
j^\msf{a}(z) := \lim_{\Delta\to1}(\Delta-1)\,O^{+\,\msf{a}}_\Delta(z,\bar z)\,.
\ee
Using the OPEs \eqref{O+O+} and \eqref{O-O+}, one can show that celestial gluons also transform as Kac-Moody primaries in the adjoint representation,
\be\label{jO}
j^\msf{a}(z_1)\,O^{\ell\,\msf{b}}_{\Delta}(z_2,\bar z_2)\sim-\frac{\im\,f^\msf{abc}}{z_{12}}\,O^{\ell\,\msf{c}}_{\Delta}(z_2,\bar z_2)\,.
\ee
Taking $\ell=+1$ and $\Delta\to1$ again in this OPE gives the usual OPE bewteen Kac-Moody currents at level $0$. Expanding the current in its holomorphic modes,
\be\label{jmodes}
j^\msf{a}(z) = \sum_{n=-\infty}^\infty\frac{j^\msf{a}_n}{z^{n+1}}\,,\qquad j^\msf{a}_n = \oint\displaylimits_0\frac{\d z}{2\pi\im}\,z^n\,j^\msf{a}(z)\,,
\ee
one can straightforwardly find their action on $O^{\ell\,\msf{a}}_\Delta$ from \eqref{jO},
\be\label{[j,O]}
[j^\msf{a}_n,O^{\ell\,\msf{b}}_\Delta(z,\bar z)] = -\im\,f^\msf{abc}\,z^n\,O^{\ell\,\msf{c}}_\Delta(z,\bar z)\,.
\ee
Including the Kac-Moody modes, we obtain the following extended algebra:
\be\label{kmalg}
[j^\msf{a}_m,j^\msf{b}_n] = -\im\,f^\msf{abc}\,j^\msf{c}_{m+n}\,,\qquad[L_m,j^\msf{a}_n] = -n\,j^\msf{a}_{m+n}\,,\qquad[\cP_{a,b},j^\msf{a}_n] = 0\,.
\ee
The last of these can be justified by taking a conformally soft limit of \eqref{PO}, or equivalently by showing that the OPE of $[\cP_{a,b},j^\msf{a}(z)]$ with an arbitrary celestial gluon is non-singular. These form the leading\footnote{We will not discuss the subleading soft gluon symmetry found in \cite{Lysov:2014csa} in this work. Since its action shifts the conformal weight by $\Delta\mapsto\Delta-1$, its ``descendants'' would enter as more --- instead of less --- singular terms in the OPE. But gluon amplitudes are clearly not infinitely singular.} global symmetry algebra of the Yang-Mills CCFT.

A typical descendant occurring as a subleading term in the OPEs \eqref{O+O+}, \eqref{O-O+} is of the form,
\be\label{desclist}
j^{\msf{a}_1}_{-k_1}\cdots j^{\msf{a}_p}_{-k_p}(L_{-1})^m(\bar L_{-1})^nO^{\ell\,\msf{c}}_\Delta(z,\bar z)\,,
\ee
with each $k_a\geq 1$. Adding these to the OPEs with unknown OPE coefficients, one can apply various symmetry generators to generate constraints on the coefficients. However, the constraints coming from conformal and Kac-Moody actions become increasingly cumbersome very quickly. But quite luckily, since $\cP_{a,b}$ commutes with the Kac-Moody generators, its action does not mix descendants of different Kac-Moody weights. In \S\ref{sec:conformal}, this allows us to use just translation invariance to fix the OPE coefficients accompanying all the purely conformal descendants $(L_{-1})^m(\bar L_{-1})^nO^{\ell\,\msf{c}}_\Delta$ without having to worry about the Kac-Moody descendants. For completeness, we also give an example of how OPE coefficients of some $j_{-1}^\msf{a}$-descendants can be determined from this data in \S\ref{sec:km}.

Note however that this idea hinges on the assumption that there are no further global symmetries of the Yang-Mills CCFT whose descendants might mix with these purely conformal descendants under translations. Thus, our calculations in \S\ref{sec:gluon} and appendix \ref{app:KM} also bolster our confidence that these are all the kinds of descendants that occur in the gluon OPE. However, more work along the lines of \cite{Banerjee:2020zlg} might be needed to confirm this.


\subsection{Conformal descendants}\label{sec:conformal}

We will use the action of $\cP_{0,-1}$ and $\cP_{-1,0}$ to determine the coefficients of conformal descendants in celestial gluon OPEs. To do this, we need the following easily derived relations,
\be\label{P0-1ac}
[\cP_{0,-1},(L_{-1})^m(\bar L_{-1})^nO^{\ell\,\msf{c}}_{\Delta_1+\Delta_2-1}(0,0)]= m\,(L_{-1})^{m-1}(\bar L_{-1})^nO^{\ell\,\msf{c}}_{\Delta_1+\Delta_2}(0,0)\,,
\ee
and
\be\label{P-10ac}
[\cP_{-1,0},(L_{-1})^m(\bar L_{-1})^nO^{\ell\,\msf{c}}_{\Delta_1+\Delta_2-1}(0,0)] = n\,(L_{-1})^m(\bar L_{-1})^{n-1}O^{\ell\,\msf{c}}_{\Delta_1+\Delta_2}(0,0)\,,
\ee
evaluated at the origin for simplicity.

Let us denote the contribution of a celestial gluon of helicity $\ell$ to the OPE of two gluons of helicity $\ell_1,\ell_2$ by
\be\label{opewilson}
O^{\ell_1\,\msf{a}}_{\Delta_1}(z_1,\bar z_1)\,O^{\ell_2\,\msf{b}}_{\Delta_2}(z_2,\bar z_2)\sim -\frac{\im\,f^\msf{abc}\,C_{\Delta_1,\Delta_2}^{\ell_1\ell_2\ell}(z_{12},\bar z_{12},\p_2,\dbar_2)\,O^{\ell\,\msf{c}}_{\Delta_1+\Delta_2-1}(z_2,\bar z_2)}{z_{12}^{h_1+h_2-h}\,\bar z_{12}^{\bar h_1+\bar h_2-\bar h}}\,,
\ee
where $(h,\bar h)$ are the conformal weights of $O^{\ell\,\msf{c}}_{\Delta_1+\Delta_2-1}$, and $C_{\Delta_1,\Delta_2}^{\ell_1\ell_2\ell}$ encodes contributions from this primary and its conformal descendants:
\be\label{Cl1l2l}
C_{\Delta_1,\Delta_2}^{\ell_1\ell_2\ell}(z_{12},\bar z_{12},\p_2,\dbar_2) = \sum_{m,n=0}^\infty C^{\ell_1\ell_2\ell}_{m,n}(\Delta_1,\Delta_2)\,z_{12}^m\,\bar z_{12}^n\,(L_{-1})^m\,(\bar L_{-1})^n\,.
\ee
Here, we are viewing $L_{-1},\bar L_{-1}$ respectively as the holomorphic and antiholomorphic derivatives $\p_2,\dbar_2$ when taken against the primary $O^{\ell\,\msf{c}}_{\Delta_1+\Delta_2-1}(z_2,\bar z_2)$.

To find recursion relations on the coefficients $C^{\ell_1\ell_2\ell}_{m,n}(\Delta_1,\Delta_2)$, we set $z_2=0=\bar z_2$ and act with $\cP_{0,-1}$ and $\cP_{-1,0}$ on \eqref{opewilson}. Applying \eqref{PO}, \eqref{P0-1ac} and \eqref{P-10ac}, this process generates
\be\label{recursion}
C^{\ell_1\ell_2\ell}_{m,n}(\Delta_1,\Delta_2) = \frac{C^{\ell_1\ell_2\ell}_{m-1,n}(\Delta_1+1,\Delta_2)}{m} = \frac{C^{\ell_1\ell_2\ell}_{m,n-1}(\Delta_1+1,\Delta_2)}{n}\,.
\ee
From the leading OPE \eqref{O+O+} and \eqref{O-O+}, we have the following boundary conditions for these recursion relations:
\be\label{bdrycond}
\begin{split}
    &C^{+\,+\,+}_{0,0}(\Delta_1,\Delta_2) = B(\Delta_1-1,\Delta_2-1)\,,\\
    &C^{-\,+\,-}_{0,0}(\Delta_1,\Delta_2) = B(\Delta_1+1,\Delta_2-1)\,,\\
    &C^{-\,+\,+}_{0,0}(\Delta_1,\Delta_2) = B(\Delta_1-1,\Delta_2+1)\,.
\end{split}
\ee
Thus, solving \eqref{recursion}, we readily discover the entire series of OPE coefficients,
\begin{align}
    &C^{+\,+\,+}_{m,n}(\Delta_1,\Delta_2) = \frac{B(\Delta_1-1,\Delta_2-1)}{m!\,n!}\,\frac{(\Delta_1-1)_{m+n}}{(\Delta_1+\Delta_2-2)_{m+n}}\,,\label{C+++}\\
    &C^{-\,+\,-}_{m,n}(\Delta_1,\Delta_2) = \frac{B(\Delta_1+1,\Delta_2-1)}{m!\,n!}\,\frac{(\Delta_1+1)_{m+n}}{(\Delta_1+\Delta_2)_{m+n}}\,,\label{C-+-}\\
    &C^{-\,+\,+}_{m,n}(\Delta_1,\Delta_2) = \frac{B(\Delta_1-1,\Delta_2+1)}{m!\,n!}\,\frac{(\Delta_1-1)_{m+n}}{(\Delta_1+\Delta_2)_{m+n}}\,,\label{C-++}
\end{align}
where $(a)_q := \Gamma(a+q)/\Gamma(a)$ are Pochhammer symbols. 

Hence, we observe that translation symmetry is a very powerful constraint on the structure of the CCFT which would be absent from a garden-variety CFT. Performing such an all order calculation using Virasoro and Kac-Moody symmetry constraints is almost inconceivable, and it is almost always more useful to work out 4-point conformal blocks rather than the descendants' OPE coefficients. However, knowing all order contributions to the OPE as we do here will help us make contact with interesting universal statements about scattering amplitudes at arbitrary multiplicity.


\subsection{Kac-Moody descendants}\label{sec:km}

As an example, we analyze the first Kac-Moody descendants contributing to the like-helicity OPE \eqref{O+O+}. In particular, we need to justify that the OPE data generated above using translation invariance is consistent with the other symmetries in the problem.

To begin with, let us take the following ansatz containing $j_{-1}^\msf{a}$-descendants:
\begin{multline}\label{O+O+km}
    O^{+\,\msf{a}}_{\Delta_1}(z_1,\bar z_1)\,O^{+\,\msf{b}}_{\Delta_2}(z_2,\bar z_2) \sim \;-\frac{\im}{z_{12}}\,B(\Delta_1-1,\Delta_2-1)\left[f^\msf{abd}\right. + f^\msf{abd}\,c_1\,z_{12}\,L_{-1}\\
    \left.+\,\im\,c_2^\msf{abcd}\,z_{12}\,j^\msf{c}_{-1} +\,\cO(\bar z_{12},z_{12}^2)\right]O^{+\,\msf{d}}_{\Delta_1+\Delta_2-1}(z_2,\bar z_2)\,,
\end{multline}
with to-be-determined coefficient functions $c_1(\Delta_1,\Delta_2)$ and $c_2^\msf{abcd}(\Delta_1,\Delta_2)$. Now, we observe that $c_1$ can be read off from \eqref{C+++} evaluated for $m=1,n=0$,
\be\label{c1}
c_1(\Delta_1,\Delta_2) = \frac{\Delta_1-1}{\Delta_1+\Delta_2-2}\,.
\ee
To fix $c_2^\msf{abcd}$, we set $z_2=0=\bar z_2$ and act with $j_1^\msf{e}$ on this OPE. Using \eqref{[j,O]} and \eqref{kmalg}, this yields the relation,
\be\label{kmeq}
f^\msf{abc}f^\msf{ecf}\,c_1 + f^\msf{ecg}f^\msf{gdf}\,c_2^\msf{abcd} = f^\msf{eac}f^\msf{cbf}\,.
\ee
To solve for the remaining coefficient, we further guess an ansatz of the form
\be\label{c2ansatz}
\qquad c_2^\msf{abcd} = \alpha\,\delta^\msf{ac}\delta^\msf{bd} + \beta\,\delta^\msf{ad}\delta^\msf{bc} + \gamma\,\delta^\msf{ab}\delta^\msf{cd}\,.
\ee
One can in principle add other group-invariant ``tensor structures'' to this ansatz, like higher degree polynomials in structure constants, but we won't need them at this level. Courtesy of the Jacobi identity, this already satisfies \eqref{kmeq} for the values,
\be\label{alphabeta}
\alpha = 1-c_1\,,\qquad\beta = c_1\,,\qquad\gamma = 0\,.
\ee
Hence, we have
\be\label{c2}
\qquad c_2^\msf{abcd}(\Delta_1,\Delta_2) = \frac{\Delta_2-1}{\Delta_1+\Delta_2-2}\,\delta^\msf{ac}\delta^\msf{bd} + \frac{\Delta_1-1}{\Delta_1+\Delta_2-2}\,\delta^\msf{ad}\delta^\msf{bc}\,.
\ee
So, we find not one but two descendants contributing at this level: $j_{-1}^\msf{a}O^\msf{b}_{\Delta_1+\Delta_2-1}$ as well as its permuted partner $j_{-1}^\msf{b}O^\msf{a}_{\Delta_1+\Delta_2-1}$.

Next, we work out the constraint coming from an application of $L_1$. The resulting relation is
\be\label{kmeqL1}
f^\msf{abe}\,(\Delta_1+\Delta_2)\,c_1 + f^\msf{cde}\,c_2^\msf{abcd} = f^\msf{abe}\,\Delta_1\,.
\ee
Substituting \eqref{c1}, \eqref{c2ansatz} into this, we find the excess condition,
\be\label{alpha-beta}
\alpha-\beta = \frac{\Delta_2-\Delta_1}{\Delta_1+\Delta_2-2}\,,
\ee
which a priori overdetermines the system of equations. But this is already beautifully satisfied by our solution \eqref{c2}. This demonstrates, at least by way of an example, that the enormous amount of symmetry in a CCFT can indeed allow for non-trivial CFT data consistent with all of it.

\medskip

For the details of how these descendants can be extracted from the OPE limit of an actual 4-gluon amplitude, the reader is directed to appendix \ref{app:KM}.


\section{Multi-gluon collinear limits}\label{sec:gluon}

The OPE of two celestial gluons was derived in \cite{Fan:2019emx} by Mellin transforming the double-collinear splitting functions. The goal of this section is to generalize this computation to obtain multi-gluon OPE from Mellin transforms of multi-collinear limits. But the former can also be computed holographically by recursively applying the 2-point OPE. Most importantly, in \S\ref{sec:n=3}, we match the contributions of conformal descendants derived in \S\ref{sec:conformal} to all orders with the triple-collinear limit. This provides a mechanism for factorization poles and residues of $4d$ amplitudes to emerge from the CCFT, generating the footprints of locality and unitarity.

A celestial amplitude of $N$ gluons, the first $n$ of which are outgoing and will be taken collinear, can be written as a CCFT correlation function,
\be\label{cAform}
\cA_N(1^{\ell_1\,\msf{a}_1}_{\Delta_1}\dots n^{\ell_n\,\msf{a}_n}_{\Delta_n}\dots N^{\ell_N\,\msf{a}_N}_{\Delta_N}) = \left\la\prod_{i=1}^n O^{\ell_i\,\msf{a}_i}_{\Delta_i}(z_i,\bar z_i)\prod_{j=n+1}^N O^{\ell_j\,\msf{a}_j,s_j}_{\Delta_j}(z_j,\bar z_j)\right\ra\,.
\ee
We can use the OPE to expand this around the multi-collinear regime. Fixing an ordering of points $|z_{12}|<|z_{23}|<\cdots<|z_{n-1,n}|$, the OPE of the collinear gluons can be accessed by sequentially applying the 2-gluon OPE. It takes the general form,
\be
\prod_{i=1}^n O^{\ell_i\,\msf{a}_i}_{\Delta_i}(z_i,\bar z_i) \sim (-\im)^{n-1}f^{\msf{a}_1\msf{a}_2\msf{b_1}}\cdots f^{\msf{b}_{n-2}\msf{a}_{n}\msf{c}}\sum_{\Delta_p,\ell}\ope\,(1^{\ell_1}\dots n^{\ell_n}\to p^\ell)\,O^{\ell\,\msf{c}}_{\Delta_p}(z_n,\bar z_n)\,,
\ee
where the quantity ``ope'' (in general a differential operator) is the celestial analog of the color-stripped splitting function in \eqref{collim} and we have suppressed its dependence on the $\Delta_i$'s. Inserting this in \eqref{cAform} leads to universal asymptotics,
\begin{multline}\label{cAfac}
\cA_N(1^{\ell_1\,\msf{a}_1}_{\Delta_1}\dots n^{\ell_n\,\msf{a}_n}_{\Delta_n}\dots N^{\ell_N\,\msf{a}_N}_{\Delta_N}) \sim (-\im)^{n-1}f^{\msf{a}_1\msf{a}_2\msf{b_1}}f^{\msf{b}_1\msf{a}_3\msf{b}_2}\cdots f^{\msf{b}_{n-2}\msf{a}_{n}\msf{c}}\\
\times\sum_{\Delta_p,\ell}\ope\,(1^{\ell_1}\dots n^{\ell_n}\to p^\ell)\,\cA_{N-n+1}(p^{\ell\,\msf{c}}_{\Delta_p}\dots N^{\ell_N\,\msf{a}_N}_{\Delta_N})\,.
\end{multline}
But \eqref{cAfac} can also be obtained from Mellin transforming \eqref{collim} (after reinstating color factors). Hence, to leading order in the collinear kinematics, holographic duality will relate the multi-gluon OPE with the splitting functions via Mellin transform,
\be\label{splope}
\prod_{i=1}^n\int_0^\infty\frac{\d\omega_i}{\omega_i}\,\omega_i^{\Delta_i}\;\spl\,(1^{\ell_1}\dots n^{\ell_n}\to p^\ell) = \ope\,(1^{\ell_1}\dots n^{\ell_n}\to p^\ell)\int_0^\infty\frac{\d\omega_p}{\omega_p}\,\omega_p^{\Delta_p}\,,
\ee
where both sides are to be viewed as ``integration kernels'' acting on the remaining non-singular momentum space amplitude. Matching these will be our primary ``test'' of celestial holography. For brevity, we will use the same notation $\ope^{(n)}_\ell(1^-\,2^-\dots k^-)$ for the ope factor that we introduced for the splitting functions in \eqref{splitdef}. Recall that this corresponds to the case when gluons $1,2,\dots,k$ among the collinear ones have negative helicity. Our focus will be on the $k=0,1$ cases, with the notation being just $\ope_\ell^{(n)}$ when $k=0$.


\subsection{$1^+\,2^+\dots n^+\to p^+$}
\label{sec:spl}

As a warm up, we consider the easiest case when all the gluons in the collinearity are of positive helicity. The descendants do not get involved at this stage, but it nonetheless gives a useful consistency check of the formalism.

First let us see what holography predicts for the multi-collinear singularity. For $n$ positive helicity gluons, the multi-gluon OPE at leading order can be calculated by using \eqref{O+O+} to perform sequential Wick contractions,
\begin{multline}\label{opecase1}
    O^{+\,\msf{a}_1}_{\Delta_1}(z_1,\bar z_1)\, O^{+\,\msf{a}_2}_{\Delta_2}(z_2,\bar z_2)\cdots O^{+\,\msf{a}_n}_{\Delta_n}(z_n,\bar z_n)\\
    \sim (-\im)^{n-1}f^{\msf{a}_1\msf{a}_2\msf{b_1}}f^{\msf{b}_1\msf{a}_3\msf{b}_2}\cdots f^{\msf{b}_{n-2}\msf{a}_{n}\msf{c}}\;\ope^{(n)}_+\; O^{+\,\msf{c}}_{\Delta_p}(z_n,\bar z_n)\,,
\end{multline}
where $\Delta_p = \sum_{i=1}^n\Delta_i - n + 1$, along with the leading order result,
\be\label{ope0}
\ope^{(n)}_+ = \frac{1}{z_{12}\,z_{23}\cdots z_{n-1,n}}\prod_{k=2}^{n}B\!\left(\sum_{i=1}^{k-1}(\Delta_i-1),\Delta_k-1\right)\,.
\ee
In $4d$ terms, this specific singularity comes from partial amplitudes in the color-order $\{1,2,\dots,n,\dots\}$. Also, in this case we did not need to keep descendants in the OPE as they always produce subleading contributions to the multi-collinear limit. For instance, keeping conformal descendants in the first contraction would generate terms with non-negative powers of $z_{12}$, etc. We will see more examples of such computations below.

Next we perform a direct calculation in the bulk. A celestial amplitude involving these gluons factorizes as
\begin{multline}\label{cAfac1}
    \cA_N (1^{+\,\msf{a}_1}_{\Delta_1} \dots N^{\ell_N\,\msf{a}_N}_{\Delta_N}) \sim (-\im)^{n-1}f^{\msf{a}_1\msf{a}_2\msf{b_1}}f^{\msf{b}_1\msf{a}_3\msf{b}_2}\cdots f^{\msf{b}_{n-2}\msf{a}_{n}\msf{c}}\\
    \times\prod_{i=1}^N\int_0^\infty\frac{\d\omega_i}{\omega_i}\,\omega_i^{\Delta_i}\;\spl^{(n)}_+\; A_{N-n+1} ( p^{+\,\msf{c}} \dots N^{\ell_N\,\msf{a}_N})\,\delta^4\!\left(\omega_p q_p+\sum_{j=n+1}^N s_j\omega_jq_j\right)\,,
\end{multline}
with $\omega_p=\sum_{i=1}^n\omega_i$ and $q_p\sim q_n$, and the relevant splitting function given in \eqref{spl0}. The Mellin integrals over $\omega_1,\dots,\omega_n$ can be easily done via a change of variables to the longitudinal-momentum fractions, $\xi_i = \omega_i / \omega_p\in(0,1)$, and the total energy $\omega_p\in(0,\infty)$. Straightforward manipulations bring \eqref{cAfac1} to the general form \eqref{cAfac}. Evaluating the left side of \eqref{splope}, one finds $\Delta_p = \sum_{i=1}^n\Delta_i-n+1$ as expected, along with the ope factor,
\be\label{Mellin1}
\begin{split}
    \ope^{(n)}_+ &= \prod_{i=1}^n \int_0^1 \frac{\d \xi_i}{\xi_i} \,\xi_i^{\Delta_i} \;\delta\! \left(1-\sum_{j=1}^n\xi_j\right)\frac{1}{\xi_n\prod_{l=1}^{n-1}\xi_l\,z_{l,l+1}}\\
    & = \frac{B(\Delta_1 - 1,\dots, \Delta_n -1)}{z_{12}\,z_{23}\cdots z_{n-1,n}} \,.
\end{split}
\ee
where $B(a_1,\dots,a_n) = \Gamma(a_1)\cdots\Gamma(a_n)/\Gamma(a_1+\cdots+a_n)$ is the multivariate beta function, and the integral has been performed by recognizing that its integrand is a standard Dirichlet distribution (\cite{kotz2004continuous}, chapter 49).\footnote{Alternatively, see appendix \ref{app:integrals} and recursively apply the beta function's defining integral formula \eqref{betaint}.} To see that this evaluation of the OPE singularity matches with \eqref{ope0}, simply rewrite the beta functions in \eqref{ope0} in terms of gamma functions. The product of gamma functions collapses telescopically, yielding a match.


\subsection{$1^-\,2^+\dots n^+\to p^\pm$}
\label{sec:spl-}

Next we come to the case when one negative and $n-1$ positive helicity gluons become collinear. Using sequential Wick contractions, this time we arrive at
\begin{multline}\label{opecase2}
    O^{-\,\msf{a}_1}_{\Delta_1}(z_1,\bar z_1)\, O^{+\,\msf{a}_2}_{\Delta_2}(z_2,\bar z_2)\cdots O^{+\,\msf{a}_n}_{\Delta_n}(z_n,\bar z_n)\sim (-\im)^{n-1}f^{\msf{a}_1\msf{a}_2\msf{b_1}}f^{\msf{b}_1\msf{a}_3\msf{b}_2}\cdots f^{\msf{b}_{n-2}\msf{a}_{n}\msf{c}}\\
    \times\left[\ope^{(n)}_-(1^-)\; O^{-\,\msf{c}}_{\Delta_p}(z_n,\bar z_n) + \ope^{(n)}_+(1^-)\; O^{+\,\msf{c}}_{\Delta_p}(z_n,\bar z_n)\right]\,,
\end{multline}
with $\Delta_p = \sum_{i=1}^n\Delta_i-n+1$ the same as before. Here, at leading order,
\be\label{ope-}
\ope^{(n)}_-(1^-) = \frac{B(\Delta_1+1,\Delta_2 - 1,\dots,\Delta_n - 1)}{z_{12}\,z_{23}\cdots z_{n-1,n}}\,,
\ee
while
\begin{multline}\label{ope+}
    \ope^{(n)}_+(1^-) = \frac{B(\Delta_1 - 1,\Delta_2+1)\,B(\sum_{l=1}^2\Delta_l - 2,\Delta_3 - 1,\dots,\Delta_n - 1)}{\bar z_{12}\,z_{23}\cdots z_{n-1,n}}\\
    + \frac{B(\Delta_1+1,\Delta_2 - 1)\,B(\sum_{l=1}^2\Delta_l - 2,\Delta_3+1)\,B(\sum_{l=1}^3\Delta_l -3,\Delta_4 -1,\dots,\Delta_n - 1)}{z_{12}\,\bar z_{23}\,z_{34}\,\cdots z_{n-1,n}}\\
    +\cdots+\frac{B(\Delta_1+1,\Delta_2 - 1,\dots,\Delta_{n-1} - 1)\,B(\sum_{l=1}^{n-1}\Delta_l - n,\Delta_n+1)}{z_{12}\cdots z_{n-2,n-1}\,\bar z_{n-1,n}}\,,\hspace{4.6em}
\end{multline}
which we have simplified by collapsing individual beta functions to multivariate beta functions. However, we will see that we will also need subleading contributions from descendants for this case.

As we did in \eqref{Mellin1}, we want to perform a Mellin transform on the splitting functions and check whether they match with the OPE. Mellin transforming $\text{split}^{(n)}_-(1^-)$ given in \eqref{spl1-}, and going to the integration variables $\omega_p,\xi_1,\dots,\xi_n$, we obtain $\Delta_p = \sum_{i=1}^n\Delta_i-n+1$ and
\be\label{Mellin2}
\begin{split}
    \ope^{(n)}_-(1^-) &= \prod_{i=1}^n \int_0^1 \frac{\d \xi_i}{\xi_i} \,\xi_i^{\Delta_i} \;\delta\! \left(1-\sum_{j=1}^n\xi_j\right)\frac{\xi_1^2}{\xi_n\prod_{l=1}^{n-1}\xi_l\,z_{l,l+1}}\\
    & = \frac{B(\Delta_1 + 1,\Delta_2-1,\dots, \Delta_n -1)}{z_{12}\,z_{23}\cdots z_{n-1,n}} \,,
\end{split}
\ee
which again matches with \eqref{ope-}.

\medskip

To venture beyond such basic consistency checks, we finally come to the Mellin transform of the second splitting function $\text{split}^{(n)}_+(1^-)$ of \eqref{spl1+}. This produces
\begin{multline}\label{splitOP}
    \text{ope}_+^{(n)}(1^-) = \prod_{i=1}^n \int_0^1 \frac{\d \xi_i}{\xi_i} \,\xi_i^{\Delta_i} \;\delta\! \left(1-\sum_{k=1}^n\xi_k\right)\frac{\xi_1^2}{\xi_n\prod_{i=1}^{n-1}\xi_i \,z_{i,i+1}}\\
    \times\left[\sum_{j=2}^{n-1}\frac{(\sum_{l=1}^j\xi_l\,z_{1l})^3}{(\sum_{l=1}^j\xi_l)\,(\sum_{l=1}^j\xi_l\,z_{jl})\,(\sum_{l=1}^j\xi_l\,z_{j+1,l})}\,\frac{z_{j,j+1}}{S_{1j}}- \frac{(\sum_{l=1}^n\xi_l\,z_{1l})^3}{(\sum_{l=1}^n\xi_l\,z_{nl})}\,\frac{1}{S_{1n}}\right]\,,
\end{multline}
where $S_{1 j} := \sum_{1\leq k<l\leq j} \xi_k \, \xi_l \, | z_{kl} |^2$. Firstly notice that this has the same number of terms as \eqref{ope+}. Each term in \eqref{splitOP} comes from a particular MHV diagram \cite{Birthwright:2005ak}, so that this counting points to a plausible 1:1 correspondence between each term in the multi-gluon OPE and an MHV diagram. We will begin with a detailed exploration of the various terms in this integral in the ``toy model'' of a triple-collinear limit. Subsequently, we will briefly explain how to scale this up to general $n$, in particular holographically recovering the $j=2$ term in the second line of \eqref{splitOP}. The means of recovering the $j\geq3$ terms and the last term are still being investigated.


\subsubsection{$n=3$}
\label{sec:n=3}

In the triple-collinear limit, we need to evaluate
\begin{multline}\label{ope3}
    \text{ope}_+^{(3)}(1^-) = \frac{1}{z_{12}\,z_{23}}\prod_{i=1}^3 \int_0^1 \frac{\d \xi_i}{\xi_i} \,\xi_i^{\Delta_i} \;\delta\! \left(1-\sum_{k=1}^3\xi_k\right)\frac{\xi_1}{\xi_2\,\xi_3}\\
    \times\left[\frac{\xi_2^3\,z_{12}^2}{\xi_1\,(\xi_1+\xi_2)\,(\xi_1\,z_{13}+\xi_2\,z_{23})}\,\frac{z_{23}}{S_{12}} + \frac{(\xi_2\,z_{12}+\xi_3\,z_{13})^3}{(\xi_1\,z_{13}+\xi_2\,z_{23})}\,\frac{1}{S_{13}}\right]\,,
\end{multline}
where $S_{12} = \xi_1\,\xi_2\,|z_{12}|^2$ and $S_{13} = \xi_1\,\xi_2\,|z_{12}|^2 + \xi_1\,\xi_3\,|z_{13}|^2 + \xi_2\,\xi_3\,|z_{23}|^2$. 

Let's look at the first term in \eqref{ope3}. Substituting for $S_{12}$ and writing $z_{13}=z_{12}+z_{23}$, it can be converted into
\begin{multline}\label{ope3t1}
    \frac{1}{\bar z_{12}\,z_{23}}\int_0^1\d\xi_1\,\xi_1^{\Delta_1-2}\int_0^1\d\xi_2\,\xi_2^{\Delta_2}\int_0^1\d\xi_3\,\xi_3^{\Delta_3-2} \;\delta\! \left(1-\sum_{i=1}^3\xi_i\right)\frac{1}{(\xi_1+\xi_2)\,(\xi_1+\xi_2-\xi_1\,z_{12}/z_{32})}\\
    = \frac{B(\Delta_1-1,\Delta_2+1)\,B(\Delta_1+\Delta_2-2,\Delta_3-1)}{\bar z_{12}\,z_{23}}\,{}_2F_1\biggl(1,\,\Delta_1-1\,;\,\Delta_1+\Delta_2\,;\,\frac{z_{12}}{z_{32}}\biggr)\,.
\end{multline}
The integration has been performed using the integral formula \eqref{2F1euler}. The beta functions are the ones we anticipated in \eqref{ope+}. But we can go even further and holographically predict the entire Gauss ${}_2F_1$ appearing in \eqref{ope3t1}. To do this, we compute the 3-gluon OPE by incorporating the contributions from conformal descendants discussed in \S\ref{sec:conformal}.

We compute the coefficient of the $1/\bar{z}_{12}\,z_{23}$ term in the 3-point OPE $O^-\,O^+\,O^+$ by using \eqref{opewilson}. This is precisely the term,
\begin{multline}\label{O-O+O+}
    O^{-\,\msf{a}}_{\Delta_1}(z_1,\bar z_1)\,O^{+\,\msf{b}}_{\Delta_2}(z_2,\bar z_2)\,O^{+\,\msf{c}}_{\Delta_3}(z_3,\bar z_3)\sim\frac{\im\,f^\msf{abd}}{\bar z_{12}}\,C^{-\,+\,+}_{\Delta_1,\Delta_2}(z_{12},\bar z_{12},\p_2,\bar\p_2)\,\frac{\im\,f^\msf{dce}}{z_{23}}\\
    \times B(\Delta_1+\Delta_2-2,\Delta_3-1)\,O^{+\,\msf{e}}_{\Delta_1+\Delta_2+\Delta_3-2}(z_3,\bar z_3) + \cdots\,.
\end{multline}
In the first Wick contraction, we keep conformal descendants as we will see that singular terms of the form $z_{12}^m/\bar{z}_{12}$, $m\geq0$, survive. We only need the leading term in the second Wick contraction because other terms would come with further descendants of $O^{+\,\msf{e}}_{\Delta_1+\Delta_2+\Delta_3-2}$ and give subleading contributions to the triple-collinear limit. Applying \eqref{Cl1l2l}, \eqref{C-++}, we see that the terms in $C^{-\,+\,+}_{\Delta_1,\Delta_2}$ that contain positive powers of $\bar\p_2$ cannot act on $1/z_{23}$,\footnote{It is natural to drop distributional terms of the form $\dbar_2\,z_{23}^{-1}\sim\delta(z_{23})$ because these can only contribute when the corresponding celestial operators exactly coincide and the leading OPE diverges. However, they might be needed for calculating subleading corrections to the multi-collinear limits. We thank Tim Adamo for pointing this out.} while the action of the rest of the terms produces the coefficient,
\begin{multline}\label{2F1series}
    C^{-\,+\,+}_{\Delta_1,\Delta_2}(z_{12},\bar z_{12},\p_2,\bar\p_2)\,\frac{1}{z_{23}} = B(\Delta_1-1,\Delta_2+1)\left(\sum_{m=0}^\infty\frac{1}{m!}\,\frac{(\Delta_1-1)_m}{(\Delta_1+\Delta_2)_m}\,z_{12}^m\,(\p_2)^m\right)\frac{1}{z_{23}}\\
    = \frac{B(\Delta_1-1,\Delta_2+1)}{z_{23}}\sum_{m=0}^\infty\frac{(\Delta_1-1)_m}{(\Delta_1+\Delta_2)_m}\left(\frac{z_{12}}{z_{32}}\right)^m\\
    = \frac{B(\Delta_1-1,\Delta_2+1)}{z_{23}}\,{}_2F_1\biggl(1,\,\Delta_1-1\,;\,\Delta_1+\Delta_2\,;\,\frac{z_{12}}{z_{32}}\biggr)\,,\hspace{2.6em}
\end{multline}
having used the series expansion given in \eqref{2F1euler}. As promised, the expected Gauss hypergeometric function is generated dynamically in the CCFT. 

\medskip

Let us also try to evaluate the second term in \eqref{ope3}, though we will only partially succeed at matching this with the OPE. The calculation of the OPE in \eqref{2F1series} motivates us to use the new variables $w := z_{12}/z_{32}$ and its conjugate as expansion parameters around the leading singularity. These allow us to reexpress the second term of \eqref{ope3} as
\begin{multline}
\label{triple}
   \frac{1}{z_{12}\,\bar z_{23}}\int_0^1\d\xi_1\,\xi_1^{\Delta_1}\int_0^1\d\xi_2\,\xi_2^{\Delta_2-2}\int_0^1\d\xi_3\,\xi_3^{\Delta_3-2}\;\delta\!\left(1-\sum_{i=1}^3\xi_i\right)\\
   \times\frac{(\xi_3-(\xi_2+\xi_3)\,w)^3}{(\xi_1+\xi_2 - \xi_1\,w)}\,\frac{1}{(\xi_1\,\xi_2\,|w|^2+\xi_1\,\xi_3\,|1-w|^2+\xi_2\,\xi_3)}\,.
\end{multline}
To manifest some of the hidden structure in this integral, we Taylor expand in $\bar w$. This results in
\begin{multline}\label{tripleexp}
    \frac{1}{z_{12}\,\bar z_{23}}\sum_{m=0}^\infty\bar w^m\int_0^1\d\xi_1\,\xi_1^{\Delta_1+m}\int_0^1\d\xi_2\,\xi_2^{\Delta_2-2}\int_0^1\d\xi_3\,\xi_3^{\Delta_3-m-3}\;\delta\!\left(1-\sum_{i=1}^3\xi_i\right)\\
    \times(\xi_3-(\xi_2+\xi_3)\,w)^{m+3}\,(\xi_1+\xi_2 - \xi_1\,w)^{-m-2}\,.
\end{multline}
Evaluating the $\xi_2$-integral, each term in this series takes the form of a type-$(3,6)$ Aomoto-Gelfand hypergeometric function (see \cite{Aomoto}, section 3.3.5),
\begin{multline}\label{ag36}
    \frac{1}{z_{12}\,\bar z_{23}}\sum_{m=0}^\infty\bar w^m\int_0^1\d\xi_1\int_0^{1-\xi_1}\d\xi_3\;\xi_1^{\Delta_1+m}\,\xi_3^{\Delta_3-m-3}\,(1-\xi_1-\xi_3)^{\Delta_2-2}\\(-w+\xi_1\,w+\xi_3)^{m+3}\,(1-\xi_1\,w-\,\xi_3)^{-m-2}\,.
\end{multline}
Such functions were already encountered in the context of celestial amplitudes in \cite{Schreiber:2017jsr}. One can perform these integrals explicitly by computing a double series expansion in $w,\bar w$ and hope to match the coefficients with the 3-gluon OPE. 

For instance at $\cO(w^0)$, setting $w=0$ while formally keeping $\bar w$ non-zero directly in \eqref{triple}, this time we arrive at a ${}_2F_1$ in $\bar w$ :
\be\label{ope3t2}
\cO(w^0)\,:\quad\!\frac{B(\Delta_1+1,\Delta_2-1)\,B(\Delta_1+\Delta_2-2,\Delta_3+1)}{z_{12}\,\bar z_{23}}\,{}_2F_1(1,\,\Delta_1+1\,;\,\Delta_1+\Delta_2\,;\,\bar w)\,.
\ee
This can be recovered by computing the coefficient of the $1/z_{12}\,\bar{z}_{23}$ term in the $O^-\,O^+\,O^+$ OPE, and again keeping only conformal descendants,
\begin{multline}\label{O-O+O+conj}
    O^{-\,\msf{a}}_{\Delta_1}(z_1,\bar z_1)\,O^{+\,\msf{b}}_{\Delta_2}(z_2,\bar z_2)\,O^{+\,\msf{c}}_{\Delta_3}(z_3,\bar z_3)\sim\frac{\im\,f^\msf{abd}}{z_{12}}\,C^{-\,+\,-}_{\Delta_1,\Delta_2}(z_{12},\bar z_{12},\p_2,\bar\p_2)\,\frac{\im\,f^\msf{dce}}{\bar z_{23}}\\
    \times B(\Delta_1+\Delta_2-2,\Delta_3+1)\,O^{+\,\msf{e}}_{\Delta_1+\Delta_2+\Delta_3-2}(z_3,\bar z_3) + \cdots\,.
\end{multline}
Using \eqref{C-+-}, we find
\begin{multline}\label{2F1conj}
    C^{-\,+\,-}_{\Delta_1,\Delta_2}(z_{12},\bar z_{12},\p_2,\bar\p_2)\,\frac{1}{\bar z_{23}} = B(\Delta_1+1,\Delta_2-1)\left(\sum_{m=0}^\infty\frac{1}{m!}\,\frac{(\Delta_1+1)_m}{(\Delta_1+\Delta_2)_m}\,\bar z_{12}^m\,(\dbar_2)^m\right)\frac{1}{\bar z_{23}}\\
    = \frac{B(\Delta_1+1,\Delta_2-1)}{\bar z_{23}}\sum_{m=0}^\infty\frac{(\Delta_1+1)_m}{(\Delta_1+\Delta_2)_m}\left(\frac{\bar z_{12}}{\bar z_{32}}\right)^m\\
    = \frac{B(\Delta_1+1,\Delta_2-1)}{\bar z_{23}}\,{}_2F_1\biggl(1,\,\Delta_1+1\,;\,\Delta_1+\Delta_2\,;\,\frac{\bar z_{12}}{\bar z_{32}}\biggr)\,.\hspace{2.6em}
\end{multline}
This agrees with \eqref{ope3t2}.

The most plausible origin of the non-trivial functional dependence of \eqref{ag36} on $w$ lies in contributions coming from Kac-Moody descendants. This will require finding an all order understanding of these analogous to our analysis of conformal descendants in \S\ref{sec:conformal}. At least at leading orders, this viewpoint is reinforced by the OPE limit of a 4-gluon amplitude derived in appendix \ref{app:KM} where such descendants do indeed show up.


\subsubsection{General $n$}
\label{sec:gen}

We can now describe some methods that may help scale up these computations in the future. 

Denote the $j^{\text{th}}$ term, $2\leq j\leq n-1$, in the expression \eqref{splitOP} by
\begin{multline}
\label{generalN}
    I_{n,j} = \prod_{i=1}^n \int_0^1 \frac{\d \xi_i}{\xi_i} \,\xi_i^{\Delta_i} \;\delta\! \left(1-\sum_{k=1}^n\xi_k\right)\frac{\xi_1^2}{\xi_n\prod_{i=1}^{n-1}\xi_i \,z_{i,i+1}}\\
    \times\frac{(\sum_{l=1}^j\xi_l\,z_{1l})^3}{(\sum_{l=1}^j\xi_l)\,(\sum_{l=1}^j\xi_l\,z_{jl})\,(\sum_{l=1}^j\xi_l\,z_{j+1,l})}\,\frac{z_{j,j+1}}{S_{1j}}\,.
\end{multline}
Observe that, except for the delta function, the integrand of $I_{n,j}$ can be factorized into a product of two functions, one depending on $\xi_1,\dots,\xi_j$ and the other on $\xi_{j+1},\dots,\xi_n$. In fact, Euler integrals like $I_{n,j}$ satisfy an elegant factorization property whereby factorization of the integrand also breaks the integral into two smaller such integrals. This is discussed in some detail in appendix \ref{app:integrals}. Using this, the Euler sub-integral involving $\xi_{j+1},\dots,\xi_n$ is found to be a simple Dirichlet integral. Then \eqref{Ifacfull} yields one of the anticipated multivariate beta functions occurring in the various terms of \eqref{ope+}, leaving us with
\begin{multline}\label{Inj}
    I_{n,j} = \frac{B(\sum_{i=1}^j\Delta_j-j,\Delta_{j+1}-1,\dots,\Delta_n-1)}{\prod_{i=1}^{n-1}z_{i,i+1}}\prod_{i=1}^{j} \int_0^1 \frac{\d \xi_i}{\xi_i} \,\xi_i^{\Delta_i}\,\delta\! \left(1-\sum_{k=1}^{j}\xi_k\right)\\
    \times\frac{\xi_1}{\xi_2\,\xi_3\cdots\xi_j}\,\frac{(\sum_{l=1}^j\xi_l\,z_{1l})^3}{(\sum_{l=1}^j\xi_l)\,(\sum_{l=1}^j\xi_l\,z_{jl})\,(\sum_{l=1}^j\xi_l\,z_{j+1,l})}\,\frac{z_{j,j+1}}{S_{1j}}\,,
\end{multline}
removing the redundancies in the calculation.

With the help of this simplification, the $j = 2$ case for instance reduces to a product of beta and hypergeometric functions,
\begin{multline} \label{hypergeo}
    I_{n,2} = \frac{B(\Delta_1 - 1,\Delta_2+1)\,B(\Delta_1+\Delta_2 - 2,\Delta_3 - 1,\dots,\Delta_n - 1)}{\bar{z}_{12}\,z_{23}\,z_{34}\cdots z_{n-1,n}}\\
    \times{}_2F_1\biggl(1,\,\Delta_1-1\,;\,\Delta_1+\Delta_2 \,;\,\frac{z_{12}}{z_{32}}\biggr)\,.
\end{multline}
This is the all multiplicity generalization of the calculation \eqref{ope3t1} of the same term in the triple-collinear case. Scaling up the Wick contractions in \eqref{O-O+O+} by adding $n-3$ further positive helicity gluon operators, and keeping conformal descendants for just the first $O^-\,O^+$ contraction, CCFT predicts precisely this singularity.

\medskip

To end this section, let us describe a systematic way to recover the rest of the leading singularities entering the expression \eqref{ope+} of $\ope^{(n)}_+(1^-)$. Inspired by the variable $w=z_{12}/z_{32}$ that showed up in the triple-collinear limit, we define a set of new variables in terms of ratios of consecutive distances,
\be
w_i := \frac{z_{i,i+1}}{z_{i+2,i+1}}\,,\qquad i \in\{1,\dots,n-2\}\,.
\ee
This choice of variables also makes sense from the standpoint of sequential Wick contractions in the multi-gluon OPE. In terms of the $w_i$, one easily derives,\footnote{Except for some boundary cases that can be handled directly.}
\be
    z_{kl} = \zeta_{kl}^j\,z_{j-1,j}\,,\qquad\zeta^j_{kl} = \left(1+\sum_{a=1}^{l-k-1}(-1)^{l-k-a}\prod_{b=a}^{l-k-1}w_{k+b-1}\right)\prod_{c=l-1}^{j-2}w_c\,.
\ee
To leading order in the ratios $w_i$,
\be
S_{1j} = |z_{j-1,j}|^2\sum_{l=1}^j\xi_l\sum_{k=1}^{l-1}\xi_k\,|\zeta^j_{kl}|^2 = \xi_j\,|z_{j-1,j}|^2\sum_{k=1}^{j-1}\xi_k + \cO(w_i)\,.
\ee
Similarly, we find the numerator factor,
\be
\left(\sum_{l=1}^j\xi_l\,z_{1l}\right)^3 = \left(z_{j-1,j}\sum_{l=1}^j\xi_l\,\zeta_{1l}^j\right)^3 = \xi_j^3\,z^3_{j-1,j} + \cO(w_i)\,,
\ee
and the denominator factors,
\begin{align}
    &\sum_{l=1}^j\xi_l\,z_{lj} = z_{j-1,j}\sum_{l=1}^{j-1}\xi_l\,\zeta^j_{lj} = z_{j-1,j}\sum_{l=1}^{j-1}\xi_l + \cO(w_i)\,,\\
    &\sum_{l=1}^j\xi_l\,z_{l,j+1} = z_{j,j+1}\sum_{l=1}^j\xi_j\,\zeta^{j+1}_{l,j+1} = z_{j,j+1}\sum_{l=1}^j\xi_j + \cO(w_i)\,.
\end{align}
The main signifance of these expansions lies in the fact that one can also systematically keep subleading terms of $\cO(w_i)$ to probe descendants exchanged in the OPE. 

With these leading order results in $w_i$, along with judicious use of the constraint $\sum_{k=1}^j\xi_k=1$, \eqref{Inj} simplifies to
\begin{multline}
\label{leadingorderint}
    I_{n,j} = \frac{B(\sum_{i=1}^j\Delta_j-j,\Delta_{j+1}-1,\dots,\Delta_n-1)}{z_{12}\cdots z_{j-2,j-1}\,\bar z_{j-1,j}\,z_{j,j+1}\cdots z_{n-1,n}}\\
    \times\int_0^1\d\xi_1\,\xi_1^{\Delta_1}\prod_{i=2}^j\int_0^1\d\xi_i\,\xi_i^{\Delta_i-2}\;\delta\!\left(1-\sum_{k=1}^j\xi_k\right)\,\frac{\xi_j^2}{(1-\xi_j)^2} + \cO(w_i)\,.
\end{multline}
Again using the factorization techniques of appendix \ref{app:integrals}, specifically \eqref{Ifac}, this evaluates to the desired leading ope singularity,
\begin{multline}
     I_{n,j} = \frac{B(\Delta_1+1,\Delta_2 - 1,\dots,\Delta_{j - 1} - 1)}{z_{12}\cdots z_{j-2,j-1}\,\bar z_{j-1,j}\,z_{j,j+1}\cdots z_{n-1,n}}\, B\!\left(\sum_{i=1}^{j-1}\Delta_i - j + 1,\Delta_j+1\right)\\ \times B\!\left(\sum_{k=1}^{j}\Delta_k - j,\Delta_{j+1} - 1,\dots,\Delta_n - 1\right) + \cO(w_i)\,.
\end{multline}
And finally, one can similarly show that the last term of \eqref{splitOP} produces the last singularity in \eqref{ope+}.


\section{Multi-graviton collinear limits}\label{sec:graviton}

One can hope to attempt similar computations for perturbative gravity. However, the collinear regime of graviton amplitudes is generically non-singular (i.e., not meromorphic). Consequently, the question of multi-collinear limits also becomes much less precise. In fact, generally it even depends on the order in which the gravitons are made collinear. Hence, in this section, we will restrict ourselves to an analysis of sequential double-collinear limits. These are the natural objects that we can expect to get mapped to sequential applications of the OPE under a Mellin transform. They would also have to act as leading order approximations to more precise notions of multi-collinear limits for sake of consistency.\footnote{In the case of multi-gluon collinear limits $\spl^{(n)}_+$ and $\spl^{(n)}_-(1^-)$, this approximation happens to yield the exact splitting functions as they only possess 2-particle factorization singularities. These also generated OPE coefficients that were permutation symmetric in the positive helicity gluons. So we did not need to discuss these issues earlier.}

\medskip

For simplicity, we will only consider the case where all helicities are positive and the case where only one helicity is negative. In the first case, on applying the 2-point OPE between two positive-helicity gravitons \eqref{grav2ope++} recursively, we have
\begin{multline}\label{gravOPEn+++}
    G_{\Delta_1}^+(z_1,\bar z_1)\,G_{\Delta_2}^+(z_2,\bar z_2)\cdots G_{\Delta_n}^+(z_n,\bar z_n) \\
    \sim\left(-\frac{\kappa}{2}\right)^{n-1}\frac{\bar z_{12}\,\bar z_{23}\cdots \bar z_{n-1,n}}{z_{12}\,z_{23}\cdots z_{n-1,n}}
    \, \prod_{j=2}^{n} B\!\left(\sum_{i=1}^{j-1}\Delta_i-1,\Delta_j-1\right)G^+_{\Delta_p}(z_n,\bar z_n)\,,
\end{multline}
where $\Delta_p = \sum_{i=1}^n{\Delta_i}$. Unlike the $n$-point OPE singularity \eqref{Mellin1} for gluons, here the coefficient does not collapse to a multivariate beta function symmetric under permutations. This means that the order in which we perform the recursive OPE is important, and from the bulk point of view it corresponds to obtaining the multi-collinear limit from repeatedly taking double-collinear limits in the same order.

In momentum space, repeated double-collinear limits give the following multi-collinear splitting function,
\begin{multline}\label{gravAllpos}
\text{split}^{(n)}_+ = \left(-\frac{\kappa}{2}\right)^{n-1}\frac{\bar{z}_{12}}{z_{12}}\, \frac{(\omega_1+\omega_2)^{2}}{\omega_{1}\, \omega_{2}}\cdot\frac{\bar{z}_{23}}{z_{23}}\, \frac{(\omega_1+\omega_2+\omega_3)^{2}}{(\omega_{1}+\omega_2)\, \omega_{3}}\cdots\frac{\bar{z}_{n-1,n}}{z_{n-1,n}}\, \frac{(\omega_1+\cdots+\omega_n)^{2}}{(\omega_{1}+\cdots+\omega_{n-1})\, \omega_{n}}\\
=\left(-\frac{\kappa}{2}\right)^{n-1}\frac{\omega_p}{\omega_1}\prod_{j=2}^{n}\frac{\sum_{i=1}^{j} \omega_i}{\omega_j}\,\frac{\bar{z}_{j-1,j}}{z_{j-1,j}} \,,\hspace{6.45cm}
\end{multline}
where $\omega_p=\sum_{i=1}^n\omega_i$, having concatenated the double-collinear splitting function $\spl^{(2)}_+$ given in \eqref{grspl} $n-1$ times. Along the lines of \eqref{splope}, Mellin transforming this gives the conformal weight $\Delta_p = \sum_{i=1}^n\Delta_i$ and the ope factor,
\be\label{ope+gr}
\begin{split}
    \ope_+^{(n)} &= \left(-\frac{\kappa}{2}\right)^{n-1}\prod_{i=1}^{n} \int_0^1 \frac{\d \xi_i}{\xi_i} \,\xi_i^{\Delta_i}\;\delta\! \left(1-\sum_{k=1}^n\xi_k\right)\frac{1}{\xi_1}\prod_{j=2}^{n}\frac{\sum_{l=1}^{j} \xi_l}{\xi_j}\,\frac{\bar{z}_{j-1,j}}{z_{j-1,j}}\\
    &= \left(-\frac{\kappa}{2}\right)^{n-1}\prod_{j=2}^{n} B\!\left(\sum_{i=1}^{j-1}\Delta_i-1,\Delta_j-1\right)\,,
\end{split}
\ee
where the integral can be performed by noticing that the integrand is an example of a generalized Dirichlet distribution (\cite{kotz2004continuous}, chapter 49). This matches the celestial CFT result \eqref{gravOPEn+++}. The non-trivial consistency check here is the fact that Mellin transforms do map concatenated splitting functions to sequential OPEs, which is a basic requirement before one can embark on an analysis of BMS descendants. 

\medskip

Similarly, the leading contribution to $\ope_-^{(n)}(1^-)$ comes from the multi-graviton OPE,
\begin{multline}\label{gravOPEn-++}
    G_{\Delta_1}^-(z_1,\bar z_1)\,G_{\Delta_2}^+(z_2,\bar z_2)\cdots G_{\Delta_n}^+(z_n,\bar z_n) \\
    \sim\left(-\frac{\kappa}{2}\right)^{n-1} \frac{\bar z_{12}\,\bar z_{23}\dots \bar z_{n-1,n}}{z_{12}\,z_{23}\cdots z_{n-1,n}}
    \, \prod_{j=2}^{n} B\left(\sum_{i=1}^{j-1}\Delta_i+3,\Delta_j-1\right)G^-_{\Delta_p}(z_n,\bar z_n) + \cdots\,,
\end{multline}
with the same weight $\Delta_p=\sum_{i=1}^n{\Delta_i}$ as before. In this case, the corresponding momentum space splitting function is found to be
\be\label{gravOneNeg}
    \text{split}^{(n)}_-(1^-)=\left(-\frac{\kappa}{2}\right)^{n-1}\frac{\omega_1^3}{\omega_p^3}\prod_{j=2}^{n}\frac{\sum_{i=1}^j\omega_i}{\omega_j}\,\frac{\bar{z}_{j-1,j}}{z_{j-1,j}} \,,
\ee
obtained from $n-1$ applications of $\spl^{(2)}_-(1^-)$ given in \eqref{grspl}.
Mellin transform then gives
\be\label{ope-gr}
\begin{split}
    \ope_-^{(n)}(1^-) &= \left(-\frac{\kappa}{2}\right)^{n-1}\prod_{i=1}^{n} \int_0^1 \frac{\d \xi_i}{\xi_i} \,\xi_i^{\Delta_i}\;\delta\! \left(1-\sum_{k=1}^n\xi_k\right)\,\xi_1^3\prod_{j=2}^{n}\frac{\sum_{l=1}^{j} \xi_l}{\xi_j}\,\frac{\bar{z}_{j-1,j}}{z_{j-1,j}}\\
    &= \left(-\frac{\kappa}{2}\right)^{n-1}\prod_{j=2}^{n} B\left(\sum_{i=1}^{j-1}\Delta_i+3,\Delta_j-1\right)\,,
\end{split}
\ee
which again matches the celestial CFT result \eqref{gravOPEn-++}.

\section{Conclusions}\label{conclusion}

The emergence of bulk physics from celestial CFT is an important subject of much ongoing research. Ideally speaking, given the OPE algebra of the holographic dual, one should be able to work out all its correlators recursively. We have shown that even in the absence of this, we can determine many interesting limits of celestial amplitudes already with the leading order OPE. Our focus has been on finding an understanding of emergent locality and unitarity through the lens of asymptotic symmetries and the celestial operator algebra. Our methods aim to utilize the operator spectrum of the CCFT to all orders, and hint at interesting organizational principles that could generate multi-particle factorization behavior from the CFT data. Moreover, they also open the doors to many interesting directions of speculation.

The operator spectrum of the Yang-Mills CCFT clearly contains much more information than we have been able to find from just translation invariance. Even though we managed to fix the contributions of all global conformal descendants and some leading examples of Kac-Moody descendants to the OPE of celestial gluons, the absence of the remaining Kac-Moody descendants is still a big gap that needs to be filled. We suspect that these extra descendants will help in finding a truly holographic derivation of all the remaining terms in the multi-collinear splitting functions discussed above. 

On a similar note, we also need to find how the subleading soft gluon symmetry of \cite{Lysov:2014csa} fits into this paradigm of primaries and descendants. This should be an interesting representation theoretic problem in its own right. In fact, the subleading soft gluon symmetry will in general impose non-trivial constraints on the multi-gluon OPE, just as it constrained the 2-gluon OPE in \cite{Pate:2019lpp}. These constraints could take the form of differential-recurrence equations like the well-known Gauss contiguous relations and give an alternative way of discovering the hypergeometric functions occurring in the Mellin transformed splitting functions. Such a method would also be easier to scale up to higher multiplicity in contrast to our technique of summing up infinite series of descendants.

Another interesting route for future work is the study of the CCFT spectrum dual to general relativity and possibly quantum gravity. Initial steps in this direction have been taken in \cite{Banerjee:2020kaa,Banerjee:2020zlg} where the OPE coefficients of some of the BMS and other descendants were computed by using symmetry constraints on the celestial graviton OPEs. However, here the set of symmetries that form a non-trivial algebra with translations is much larger, obstructing an all order computation analogous to that in \S\ref{sec:conformal}. But the recent work on a double copy for celestial amplitudes \cite{Casali:2020vuy} holds promise to overcome these issues. It should be possible to find a notion of color-kinematics duality that acts as an algebra homomorphism on the CCFT operator algebra and maps celestial gluon OPEs to those of celestial gravitons. Hints of this are already present in our results from \S\ref{sec:km}. There, if one maps $j^\msf{a}_{-1} O^\msf{b}_{\Delta_1+\Delta_2-1}$ and $j^\msf{b}_{-1} O^\msf{a}_{\Delta_1+\Delta_2-1}$ to $\cP_{-2,-1}G^+_{\Delta_1+\Delta_2-1}$ and $-\cP_{-2,-1}G^+_{\Delta_1+\Delta_2-1}$ respectively in the notation of \cite{Banerjee:2020kaa}, then the OPE coefficient for the graviton supertranslation descendant $\cP_{-2,-1}G^+_{\Delta_1+\Delta_2-1}$ is found to be $\alpha-\beta$. This is given in \eqref{alpha-beta} and matches with the result found in section 8.3 of \cite{Banerjee:2020kaa}. Also, the vanishing of the OPE coefficient of its antiholomorphic partner $\cP_{-1,-2}G^+_{\Delta_1+\Delta_2-1}$ seems to go hand in hand with the absence of antiholomorphic Kac-Moody descendants in the $O^+\,O^+$ gluon OPE. Similar lines of research could also be explored in the full Einstein-Yang-Mills theory.

Above, we also saw that we only possess limited knowledge of the multi-collinear behavior of gravity. Celestial CFT might also be able to help with this by giving a concrete foundation for the gravitational MHV formalism of \cite{BjerrumBohr:2005jr}. We saw indications of this when working out the multi-gluon OPEs in \S\ref{sec:gluon}. There, various terms in the multi-collinear splitting functions were in 1:1 correspondence with both terms in the CSW recursion relations and the multi-gluon OPE singularities. This might have a straightforward generalization to gravity and help in making novel universal statements.

Finally, we would like to mention that one of our original hopes in deriving subleading terms in the OPEs was to find conformal block expansions for 4-point celestial amplitudes. There has been some work on partial wave expansions in \cite{Nandan:2019jas,Law:2020xcf}, but relating them to the operators flowing in the celestial OPE is still an open question of great import. This will bring us a step closer to viewing scattering amplitudes as conformal correlators.

\acknowledgments 
We thank Tim Adamo, Zvi Bern, Gary Horowitz and Michael Zlotnikov for useful discussions. S.E. acknowledges funding from the Bhaumik Institute. A.S. is supported by a Mathematical Institute Studentship, Oxford. D.W. is supported by NSF grant PHY1801805.

\begin{appendix}

\section{Descendants from 4-gluon amplitude}\label{app:KM}

In this appendix, we look at the 4-point gluon amplitude in the OPE limit. We follow an analogous calculation for the 4-graviton amplitude carried out in \cite{Banerjee:2020kaa}. Since we need to work with 3-point amplitudes, we will go to split signature for the sake of this section. This entails using a reality condition with $z_i$ and $\bar z_i$ being independent real variables.

The 4-point gluon amplitude with gluons $1, 2$ negative helicity and gluons $3, 4$ positive helicity is given by
\be\label{fullA4}
A_4(1^{-\,\msf{a}}\,2^{-\,\msf{b}}\,3^{+\,\msf{c}}\,4^{+\,\msf{d}}) = f^\msf{ade}f^\msf{cbe}\,\frac{\la1\,2\ra^3}{\la2\,3\ra\,\la3\,4\ra\,\la4\,1\ra} + f^\msf{ace}f^\msf{dbe}\,\frac{\la1\,2\ra^3}{\la2\,4\ra\,\la4\,3\ra\,\la3\,1\ra}\,.
\ee
We take the negative helicity gluons to be incoming and the positive helicity ones outgoing. Then the momentum conserving delta function can be expressed as follows:
\be\label{mom}
\begin{split}
    &\delta^4(\omega_1q_1+\omega_2q_2-\omega_3q_3-\omega_4q_4)\\
    &=\frac{1}{\omega_1^*\,\omega_2^*\,z_{12}^2}\,\delta(\omega_1-\omega_1^*)\,\delta(\omega_2-\omega_2^*)\,\delta\!\left(\bar z_{14}+\frac{\omega_3}{\omega_1^*}\,\frac{z_{23}}{z_{12}}\,\bar z_{34}\right)\delta\!\left(\bar z_{24}-\frac{\omega_3}{\omega_2^*}\,\frac{z_{13}}{z_{12}}\,\bar z_{34}\right)\,,
\end{split}
\ee
where
\be\label{omegasol}
\omega_1^* = \frac{z_{42}}{z_{12}}\,(\omega_3+\omega_4)+\frac{z_{34}}{z_{12}}\,\omega_3\,,\qquad\omega_2^* = \frac{z_{14}}{z_{12}}\,(\omega_3+\omega_4)-\frac{z_{34}}{z_{12}}\,\omega_3\,.
\ee
To do the Mellin transform, we make the substitutions,
\be\label{subs}
\omega_3 = t\,\omega\,,\qquad\omega_4 = (1-t)\,\omega\,,
\ee
and integrate over $\omega\in\R^+$ and $t\in(0,1)$. 

In the OPE limit $z_{34}\to0$, this produces the following asymptotics for the celestial amplitude:
\begin{multline}\label{4opelim}
    \cA_4(1^{-\,\msf{a}}_{\Delta_1}\,2_{\Delta_2}^{-\,\msf{b}}\,3_{\Delta_3}^{+\,\msf{c}}\,4_{\Delta_4}^{+\,\msf{d}})\sim \frac{1}{z_{34}}\left[f^\msf{ace}f^\msf{bde}\,\biggl(1-\frac{z_{34}}{z_{14}}\biggr)^{-1} + f^\msf{dae}f^\msf{bce}\,\biggl(1-\frac{z_{34}}{z_{24}}\biggr)^{-1}\right]\\
    \times\left\{\int_0^1\d t\,t^{\Delta_3-2}\,(1-t)^{\Delta_4-2}\left(1-\frac{z_{34}}{z_{24}}\,t\right)^{\Delta_1-1}\left(1-\frac{z_{34}}{z_{14}}\,t\right)^{\Delta_2-1} + \cO(\bar z_{34})\right\}\\
    \times\widetilde\cA_3(1_{\Delta_1}^-\,2_{\Delta_2}^-\,4^+_{\Delta_3+\Delta_4-1})\,.
\end{multline}
The $\cO(\bar z_{34})$ term comes from Taylor expanding the last two delta function factors in \eqref{mom} and will contain contributions from $\bar L_{-1}$-descendants which can be determined analogously to our procedure below. For illustration purposes, we will only focus on the $L_{-1}$ and Kac-Moody descendants. Also, here $\widetilde\cA_3$ denotes the color-stripped 3-point MHV celestial amplitude \cite{Pasterski:2017ylz},
\be\label{3opelim}
\widetilde\cA_3(1_{\Delta_1}^-\,2_{\Delta_2}^-\,4^+_{\Delta_3+\Delta_4-1}) = \Theta\!\left(\frac{z_{42}}{z_{12}}\right)\Theta\!\left(\frac{z_{14}}{z_{12}}\right)\,\frac{2\pi\,\delta(\im\,(4-\sum_{j=1}^4\Delta_j))\,\delta(\bar z_{14})\,\delta(\bar z_{24})}{z_{12}^{1-\Delta_3-\Delta_4}z_{14}^{2-\Delta_2}z_{42}^{2-\Delta_1}}\,,
\ee
(with $\Delta_j\in1+\im\,\R$ assumed as usual). The leftover $t$-integral is an Appell's $F_1$ hypergeometric function, with series expansion
\begin{multline}
    \int_0^1\d t\,t^{\Delta_3-2}\,(1-t)^{\Delta_4-2}\left(1-\frac{z_{34}}{z_{24}}\,t\right)^{\Delta_1-1}\left(1-\frac{z_{34}}{z_{14}}\,t\right)^{\Delta_2-1}\\
    = B(\Delta_3-1,\Delta_4-1)\sum_{m,n=0}^\infty\frac{(\Delta_3-1)_{m+n}\,(1-\Delta_1)_m\,(1-\Delta_2)_n}{m!\,n!\,(\Delta_3+\Delta_4-2)_{m+n}}\left(\frac{z_{34}}{z_{14}}\right)^n\left(\frac{z_{34}}{z_{24}}\right)^m\,.
\end{multline}
We will just keep the $m+n=0,1$ terms for simplicity.

Expanding \eqref{4opelim} to $\cO(z_{34}^1\,\bar z_{34}^0)$, some simple manipulations yield
\begin{multline}\label{A4asym}
    \cA_4(1^{-\,\msf{a}}_{\Delta_1}\,2_{\Delta_2}^{-\,\msf{b}}\,3_{\Delta_3}^{+\,\msf{c}}\,4_{\Delta_4}^{+\,\msf{d}})\sim\frac{B(\Delta_3-1,\Delta_4-1)}{z_{34}}\left[f^\msf{abe}f^\msf{cde}\left(1+\frac{\Delta_3-1}{\Delta_3+\Delta_4-2}\,z_{34}\,\p_4\right)\right.\\ \left. + \frac{\Delta_4-1}{\Delta_3+\Delta_4-2}\,z_{34}\left(\frac{f^\msf{ace}f^\msf{bde}}{z_{14}} + \frac{f^\msf{dae}f^\msf{bce}}{z_{24}}\right) - \frac{\Delta_3-1}{\Delta_3+\Delta_4-2}\,z_{34}\left(\frac{f^\msf{ace}f^\msf{bde}}{z_{24}} + \frac{f^\msf{dae}f^\msf{bce}}{z_{14}}\right)\right.\\
    \left.+\,\cO(z_{34}^2,\bar z_{34})\right]\widetilde\cA_3(1_{\Delta_1}^-\,2_{\Delta_2}^-\,4^+_{\Delta_3+\Delta_4-1})\,.
\end{multline}
From this and the Virasoro and Kac-Moody Ward identities, one reads off the descendants and their OPE coefficients. See for instance \cite{Blumenhagen:2009zz}, section 3.5 for an introduction to computing correlators of Kac-Moody descendants via the OPE and the global residue theorem. As expected, they can be made to agree with the coefficients predicted from symmetry in \eqref{c1}, \eqref{c2} by replacing particle labels 1 and 2 with 3 and 4 respectively.

\section{Factorization of Euler integrals}\label{app:integrals}

We first note some useful integral representations of special functions. The Euler beta function is given by
\be\label{betaint}
B(a,b) = \int_0^1\d\xi\;\xi^{a-1}(1-\xi)^{b-1} = \frac{\Gamma(a)\,\Gamma(b)}{\Gamma(a+b)}\,.
\ee
The Gauss hypergeometric function can be represented by a similar Euler integral formula as well as by a series expansion,
\be\label{2F1euler}
{}_2F_1(a,\,b\,;\,c\,;\,z) = \frac{1}{B(b,c-b)}\int_0^1\d\xi\;\xi^{b-1}(1-\xi)^{c-b-1}(1-\xi\,z)^{-a} = \sum_{n=0}^\infty\frac{(a)_n(b)_n}{(c)_n}\frac{z^n}{n!}\,.
\ee
Such integrals occur numerous times in \S\ref{sec:gluon}.

Next, we provide some standard tricks that can be applied to recursively simplify the Euler-type integrals occurring in this work. Suppose we start with an integral of the form
\be\label{eulerint}
\mathfrak{I}[f\cdot g] = \prod_{i=1}^n\int_0^1\frac{\d\xi_i}{\xi_i}\;\delta\!\left(1-\sum_{j=1}^n\xi_j\right)f(\xi_1,\dots,\xi_k)\,g(\xi_{k+1},\dots,\xi_n)\,,
\ee
for a pair of integrable functions $f$ and $g$. Also assume that $g$ is a homogeneous function of degree $\beta$ under a diagonal rescaling,
\be\label{ghom}
g(t\,\xi_{k+1},\dots,t\,\xi_n) = t^\beta g(\xi_{k+1},\dots,\xi_n)\,,\qquad t\in\R^*\,.
\ee
This will be a ubiquitous property in our splitting functions. Since the dependence on $\xi_i$'s is factorized in the integrand, we now show that this is also enough to factorize the integral into two smaller Euler integrals.

Simply insert identity in the form,
\be\label{1}
1 = \int_0^1\frac{\d\xi_0}{\xi_0}\;\delta\!\left(1-\sum_{i=k+1}^n\frac{\xi_i}{\xi_0}\right)\,.
\ee
Since $0<\sum_{i=k+1}^n\xi_i<1$ due to the delta function constraint in \eqref{eulerint}, we only need to integrate $\xi_0$ over this range. Inserting this in $\mathfrak{I}[f\cdot g]$ and rescaling $\xi_i\mapsto\xi_0\,\xi_i$ for $i=k+1,\dots,n$ produces
\be\label{Ifac}
\begin{split}
    \mathfrak{I}[f\cdot g] &= \prod_{i=0}^n\int_0^1\frac{\d\xi_i}{\xi_i}\;\delta\!\left(1-\sum_{j=0}^k\xi_j\right)\delta\!\left(1-\sum_{l=k+1}^n\frac{\xi_l}{\xi_0}\right)f(\xi_1,\dots,\xi_k)\,g(\xi_{k+1},\dots,\xi_n)\,,\\
    &= \mathfrak{I}[\xi_0^\beta f]\,\mathfrak{I}[g]\,,
\end{split}
\ee
where the sub-integrals are given by
\be\label{Ifactors}
\begin{split}
    &\mathfrak{I}[\xi_0^\beta f] = \prod_{i=0}^k\int_0^1\frac{\d\xi_i}{\xi_i}\,\delta\!\left(1-\sum_{j=0}^k\xi_j\right)\xi_0^\beta f(\xi_1,\dots,\xi_k)\,,\\
    &\mathfrak{I}[g] = \prod_{i=k+1}^n\int_0^1\frac{\d\xi_i}{\xi_i}\,\delta\!\left(1-\sum_{j=k+1}^n\xi_j\right)g(\xi_{k+1},\dots,\xi_n)\,.
\end{split}
\ee
Moreover, if $f$ is also a homogeneous function of degree $\alpha$ under a diagonal rescaling, say
\be\label{fhom}
f(t\,\xi_1,\dots,t\,\xi_k) = t^\alpha f(\xi_1\dots,\xi_k)\,,\qquad t\in\R^*\,,
\ee
we can further factorize $\mathfrak{I}[\xi_0^\beta f]$ along similar lines. The final result is
\be\label{Ifacfull}
\mathfrak{I}[f\cdot g] = B(\alpha,\beta)\,\mathfrak{I}[f]\,\mathfrak{I}[g]\,,
\ee
where, as expected,
\be\label{Ifactors2}
\mathfrak{I}[f] = \prod_{i=1}^k\int_0^1\frac{\d\xi_i}{\xi_i}\,\delta\!\left(1-\sum_{j=1}^k\xi_j\right)f(\xi_1,\dots,\xi_k)\,.
\ee
This factorization will be very useful in extracting beta functions from complicated integrals. Such properties may also be helpful in studying more general factorization behaviors of celestial amplitudes in the future.

\end{appendix}

\bibliographystyle{JHEP}
\bibliography{multicol}

\end{document}